\documentclass[letterpaper, 10 pt, conference]{ieeeconf}  
\usepackage[utf8]{inputenc}
\usepackage[T1]{fontenc}
\usepackage{amsmath}
\usepackage{graphicx}
\usepackage{breqn}
\usepackage{booktabs}
\usepackage{hyperref}

\IEEEoverridecommandlockouts                              

\usepackage{graphics} 
\usepackage{epsfig} 
\usepackage{mathptmx} 
\usepackage{amsmath} 
\usepackage{amssymb}  

\title{\LARGE \bf
A Deep Learning Based Multitask Network for Respiration Rate Estimation - A Practical Perspective
}

\author{Kapil Singh Rathore$^{1,2*}$, Sricharan Vijayarangan$^{1,2*}$, Preejith SP$^{1}$,\\ Mohanasankar Sivaprakasam$^{1,2}$  
\thanks{* Equal Contribution}
\thanks{$^{1}$ are with Healthcare Technology and Innovation Center (HTIC),
        Indian Institute of Technology (IITM), India
        {\tt\small sricharanv@htic.iitm.ac.in}}%
\thanks{$^{2}$ are with Department of Electrical Engineering,
        Indian Institute of Technology, Madras (IITM), India
        {}}%
}

\begin{document}
\maketitle
\thispagestyle{empty}
\pagestyle{empty}
\begin{abstract}
The exponential rise in wearable sensors has garnered significant interest in assessing the physiological parameters during day-to-day activities. Respiration rate is one of the vital parameters used in the performance assessment of lifestyle activities. However, obtrusive setup for measurement, motion artifacts, and other noises complicate the process. 
This paper presents a multitasking architecture based on Deep Learning (DL) for estimating instantaneous and average respiration rate from ECG and accelerometer signals, such that it performs efficiently under daily living activities like cycling, walking, etc. The multitasking network consists of a combination of Encoder-Decoder and Encoder-IncResNet, to fetch the average respiration rate and the respiration signal. The respiration signal can be leveraged to obtain the breathing peaks and instantaneous breathing cycles. Mean absolute error (MAE), Root mean square error (RMSE), inference time, and parameter count analysis has been used to compare the network with the current state of art Machine Learning (ML) model and other DL models developed in previous studies. Other DL configurations based on a variety of inputs are also developed as a part of the work. The proposed model showed better overall accuracy and gave better results than individual modalities during different activities.

\end{abstract}

\section{INTRODUCTION}

As the modern lifestyle of human beings is becoming hectic, it is imperative to estimate and monitor physiological parameters efficiently. Respiration Rate (RR) is one of the most critical parameters used for assessing the health conditions and performance during different activities. Given the current challenges presented by COVID-19, several research studies like \cite{miller2020analyzing} have emphasized the importance of monitoring the respiration rate in point of care settings. Classical methods to estimate RR are pretty reliable but require an obtrusive setup for measurement. On the other hand, physiological signals like Electrocardiogram (ECG) and Photoplethysmography (PPG) are modulated by respiration and can extract RR. 

A detailed review of techniques to extract RR from ECG and PPG signal is presented in \cite{charlton2017breathing}.
Apart from ECG and PPG, accelerometer signals to extract the respiration signal and RR are also explored like the method based on Principal Component Analysis (PCA) is presented in \cite{liu2011estimation}.
However, motion artifacts and other noises like Mayer waves makes these techniques error prone. These problems were addressed by \cite{charlton2017breathing} through the concept of fusion, which compensates for the inherent limitation in each method and enhances RR estimation accuracy.   

Of late, Machine Learning (ML) and Deep Learning (DL) algorithms are gaining much traction and are employed in multiple studies. \textit{Birrenkott et al.} \cite{birrenkott2016robust} proposed to fuse the respiration rates obtained from different modulations of ECG and PPG using Machine Learning (ML) models to determine weights for each modality based on Respiratory Quality Indices (RQI). \textit{Bian et al.} \cite{bian2020respiratory} proposed a Convolutional Neural Network (CNN) to obtain the respiration rate from PPG waveforms.  These studies were focused on estimating the average respiration rate for a window of fixed size. This problem formulation does not allow the estimation of individual respiration peak locations, which is required to estimate instantaneous respiration rate. This limitation was addressed in \cite{ravichandran2019respnet} which proposed a RespNet (RN) to extract the respiration signal from PPG. However, this formulation required a considerable jump in run time, rendering it ineffective in a point of care settings. Moreover, all these works were directed at clinical data, and their efficacy on ambulatory datasets was not studied.
 
We aim to obtain robust RR estimates using ECG and accelerometer signals derived from a chest-worn sensor while performing specific activities (walking, cycling, etc.), simultaneously optimizing it for inference speed. To this end, we propose a multitasking network that gives out both the average respiration rate and the instantaneous breathing cycles and leverages prior respiration estimates. The multitasking formulation ensures that the model optimizes actual breathing peak location while simultaneously predicting accurate breathing rates, enabling overall output improvement. Additionally, we extensively compare the model against other works and several other DL configurations. Furthermore, we have thoroughly evaluated the proposed model during various activities, using different plots.  

\section{Methodology}
\subsection{\textbf{Data Set Description}}
This study utilizes the PPG field study dataset \cite{reiss2019deep} which comprises of raw ECG, accelerometer, and respiration signal waveforms obtained from a chest-worn device. All the three signals were recorded at a sampling frequency of 700 Hz. The dataset also contains reference instantaneous heart rate for an 8-second window stridden by 2 seconds along with the corresponding R peak locations. 

Fifteen participants, seven males and eight females, were included in the study. The data collection protocol comprised of eight different activities: sitting, table soccer, cycling, driving, lunch break, walking, working, ascending and descending stairs. 

Out of the 15 subjects, subject S6 is discarded from the study due to improper recording.

\subsection{\textbf{Respiration Signal and Respiration Rate Extraction}}
\label{section:resp}
The variation in R-R interval (RRint) and R peak amplitudes (Rpeak) were used to extract the respiration signal from ECG as given in \cite{7513278}. The respiration signals were extracted for 32 second windows from the extracted R peak amplitude and R-R interval. Accelerometer-based respiration signal (ADR) was obtained using a process described in \cite{hung2008estimation}. RR from the obtained respiration signals and reference respiration signal was derived using an advanced counting algorithm \cite{schafer2008estimation}. Based on the breathing peak locations, the duration between adjacent peaks and the duration between the first and last peak of the 32 second window can used to estimate the instantaneous and average RR respectively.

\begin{figure}[t]
    \centering
    \includegraphics[width=1\linewidth,height = 0.4\linewidth]{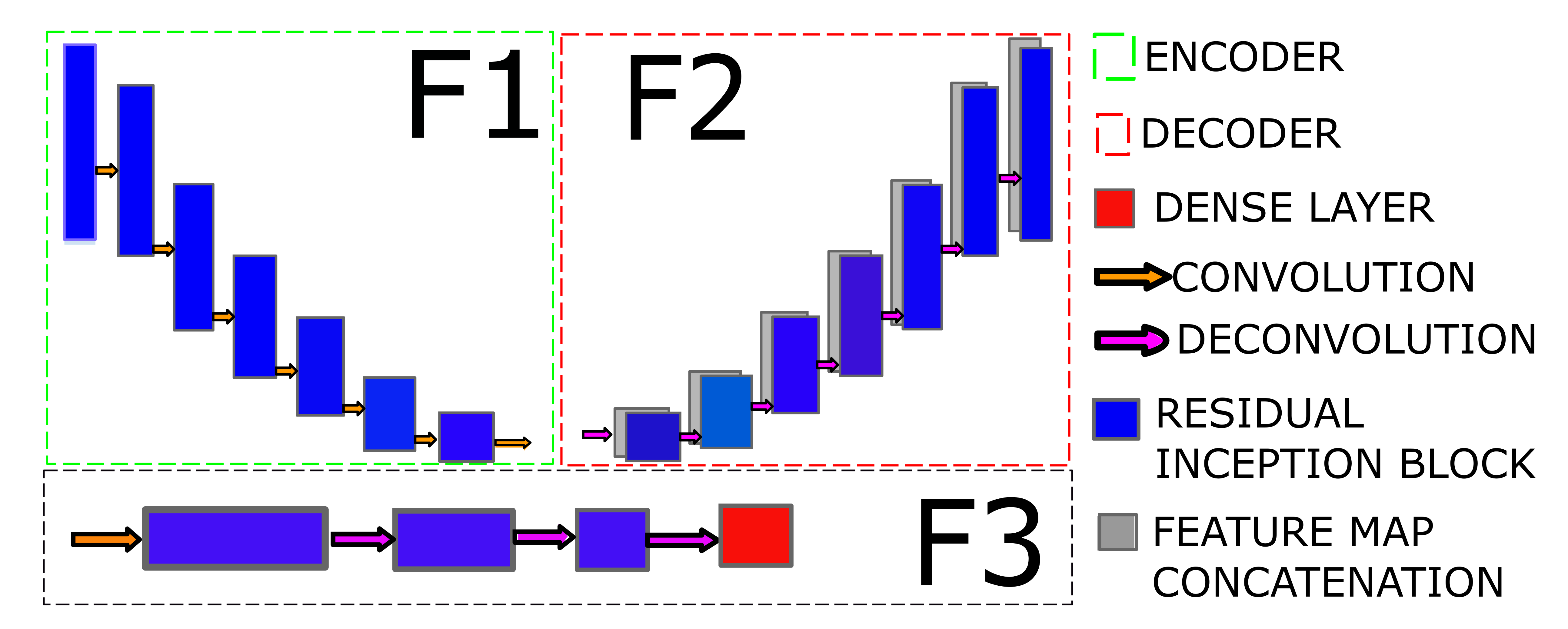}
    \caption{Description of blocks used to make DL architecture \textbf{$F_1$} represents Encoder block, \textbf{$F_2$} represents Decoder block, \textbf{$F_3$} is down sampling block with dense layer.}
    \label{fig:Architecture}
    \vspace{-1.2em}
\end{figure}

\subsection{\textbf{Model Architecture and Configurations}}
The inherent design of the architecture was inspired from \cite{ravichandran2019respnet}. However, we fed the intermediate respiration signals as input to the model for adequate optimization of inference time rather than adopting an end-to-end DL formulation. The intermediate respiration signals were extracted using the methods given in Section \ref{section:resp}. Additionally, we used an IncResNet block followed by a dense layer to introduce the multitasking functionality. The 1D convolution layer of the IncResNet block has a kernel size 4 and stride 2.

The number of filters in the Convolution layer was reduced by two, starting from 128. Furthermore a single unit dense layer was added to get the average RR.
For the encoder, the 1D convolution with kernel size 3, stride 2 was used for downsampling. The filter count was advanced by two from 32 to 1024. 
For the decoder, the upsampling is performed by 1D transposed convolution of kernel size 3 and stride 2, and the filter count was reduced by two starting from 512 to get the output respiration signal. The 1D convolution layer of the encoder, IncresNet, and the 1D convolution transpose layer of decoder were followed by Batch Normalization, Leaky ReLU activation with slope 0.2, and Inception-Res block. The kernel size of the Inception-Res block was fixed to 15 empirically with a stride of 1.

Multiple DL architectures compatible with different types of input signals were developed as part of this work, which henceforth is referred to as configurations (\textbf{CONF}). All the configurations, including the proposed multitask network (\textbf{CONF-E}), employ either a fully convolutional Encoder-Decoder, an Encoder-IncResNet, or both. A brief description of input, output, and architecture combination used for each configuration is given in Figure \ref{fig:new_model_config}.

\subsection{\textbf{DL Problem Formulation}}
The problem statement for the proposed model involves feeding in the respiration, raw ECG and accelerometer signals as input to obtain the point estimate of the respiration signal and the overall respiration rate
The inputs $X = { (x^{(11)}, x^{(12)}, x^{(13)},y^{(11)}, y^{(12)}), (x^{(21)}, x^{(22)}, x^{(23)},y^{(21)},}$
${ y^{(22)}),.... ,(x^{(n1)}, x^{(n2)}, x^{(n3)},y^{(n1)}, y^{(n2)}) }$ were fed into the DL model to obtain $\hat{y}^{(n1)}$, a point respiration rate and $\hat{y}^{(n2)}$, the respiration signal waveform where ($x^{(i)},y^{(i)} \in R^{n}$). 

For configuration C, D and E, the encoder ($F_{1}$) takes the respiration waveforms obtained from the three modalities as input $x^{(i)}$ and subsequently downsamples it multiple times to produce a compressed feature vector $Z^{(i)}$. For all the other configurations, the inputs are the raw ECG and acceleration waveform. For configuration B, C and E, the decoder ($F_{2}$) upsamples the feature vector obtained from the bottleneck layer to produce the respiration waveform $y_{pred_1}$. The average respiration rate for configuration A, B, D and E was obtained through further downsampling of the bottleneck $Z^{(i)}$ by IncRes units  followed by a dense layer ($F_{3}$). Equations \eqref{eq_new} and \eqref{eq_new2} further elucidate the above architecture for the proposed network CONF-E, 
\begin{equation}
z^{(i)} = F_{1}(x^{(i)};\theta1)    
\label{eq_new}
\end{equation}
\begin{equation}
\hat{y}^{(i1)} = F_{2}(z^{(i)};\theta2)
\label{eq_new2}
\end{equation}
\begin{equation}
\hat{y}^{(i2)} = F_{3}(z^{(i)};\theta3)   
\end{equation}

where $F_{1}$, $F_{2}$ and $F_{3}$ are the representations of the encoder, decoder and combination of Incres units and dense layer with parameters $\theta_{1}$, $\theta_{2}$ and $\theta_{3}$ respectively. The weights and biases in the proposed architecture were optimized by minimizing the Smooth$L_{1}$ loss between $\hat{y}^{(ni)}$ and $y^{(ni)}$. The Loss function L(X) is defined as: 

\begin{equation}
L_1(X) = \sum_{i=1}^{m} SmoothL_1(y_{diff_1})
\end{equation}
\begin{equation}
L_2(X) = \sum_{i=1}^{m} SmoothL_1(y_{diff_2})
\end{equation}
\begin{equation}
y_{diff_1} = y^{(i1)} - \hat{y}^{(i1)}
\end{equation}
\begin{equation}
y_{diff_2} = y^{(i2)} - \hat{y}^{(i2)}
\end{equation}

\begin{equation}
SmoothL_{1}(y_{diffi}) =  \begin{cases} 0.5(y_{diffi}^{2}), & abs(y_{diffi})< 1 \mbox{} \\ abs(y_{diffi}) - 0.5, &\mbox{otherwise,} \end{cases}
\label{eq1}
\end{equation}

\subsection{\textbf{Experimental and Evaluation details}}
The training and test dataset was taken in the ratio $80:20$.
The parameters of the DL model were randomly initialized during training. 
Adam optimizer and the SmoothL1 loss function has been utilized as this loss possesses the characteristics of both Mean Square Error (MSE) and Mean Absolute Error (MAE) when the absolute value of the argument varies from low to high as given in $eq.\eqref{eq1}$. 
 
Each configuration was trained for 100 epochs. The learning rate (LR) for CONF-C was $0.0001$, while for other configurations, an adaptive learning rate was used as given in $eq.\eqref{eq2}$. Each 32 second window of raw signal input was downsampled from 22400 to 2048 samples, therefore CONF-A, and CONF-B takes the input shape of (814,2048,3). The input respiration signal was downsampled to 4Hz as given in Section \ref{section:resp} and therefore CONF-C, CONF-D and CONF-E takes the input shape of (814,128,3). The input batch size for each configuration is kept as 128.

The ML based smart fusion (SF) \cite{birrenkott2016robust}, the CNN based model \cite{bian2020respiratory} and RN \cite{ravichandran2019respnet} are also developed for comparative purpose. The design specification for these models are adapted from their respective literature. 
The model was implemented in Tensorflow 
on a workstation housing an Nvidia GTX1080Ti 11GB GPU. 

\begin{equation}
LR =  \begin{cases} $0.01$, & no. of epochs<= 20 \mbox{} \\ $0.0001$, &\mbox{otherwise} \end{cases}
\label{eq2}
\end{equation}

Evaluation of average respiration rates was done by obtaining the Mean Absolute Error (MAE) and Root Mean Square Error (RMSE) between the obtained average RR and the ground truth RR. For configurations that provides the respiration waveform as output, the individual breathing cycles were obtained, after which the RMSE and MAE between the RR and the reference instantaneous RR were computed as given in Section \ref{section:resp}.  

\begin{table}
\centering
\caption{COMPARISON OF PROPOSED MODEL WITH DIFFERENT DL MODELS AND OTHER DL CONFIGURATIONS DEVELOPED}
\label{dl_model}
\begin{tabular}{|l|l|l|l|l|l|l|} \hline
 & \multicolumn{2}{l|}{Avg RR} & \multicolumn{2}{l|}{Inst RR} & PC(M) & Time
 \\ \hline
 & MAE & RMSE & MAE & RMSE &  & (in ms) \\ \hline
SF \cite{birrenkott2016robust} & 2.94 & 3.88 & - & - & 1.8  & 0.36  \\ \hline
CNN \cite{bian2020respiratory} & 3.55 & 4.38 & - & - &0.457  &4.20  \\ \hline
RN \cite{ravichandran2019respnet} & 3.10 & 3.93 &3.36 &4.29  &39.16  &25.99\\ \hline
CONF-A & 2.54 & 3.15 & - & - &16.41  &35.15  \\ \hline
CONF-B & 2.46 & 3.08 &3.07  &3.83  &49.37  &39.94 \\ \hline
CONF-C & 2.94 & 3.77 & 3.02 & 3.83 &20.85  &5.56  \\ \hline
CONF-D & 2.45 & 3.06 &-  &-  &16.64  &3.62 \\ \hline
\textbf{CONF-E} & \textbf{2.35} & \textbf{2.95} & \textbf{2.95} & \textbf{3.69} &\textbf{23.04}  &\textbf{9.47}  \\ \hline
\end{tabular}
\vspace{-1.5em}
\end{table}

\section{Results and Discussion}
\subsection{\textbf{Comparison of Multitasking Model with other ML and DL Configurations}}
In recent research studies, DL models have almost consistently outperformed ML models, especially with more data. However, the model's parameter count and inference time should be optimum to operate it in point of care settings. Hence, it is essential to offer insights on parameter count and run time apart from error scores. To this end, Table \ref{dl_model} consolidates the results of the comparison between the proposed multitask model (CONF-E), the other DL configurations, SF, CNN, and RN, based on MAE, RMSE, time taken (milliseconds), and parameter count (PC) (in a million). Among all configurations, the proposed multitask network (CONF-E) provides the lowest MAE and RMSE for average RR. It performs marginally better than CONF-C in terms of error scores on instantaneous RR, but the same was significantly less compared to other models for instantaneous RR. 
While our model has taken a higher parameter count and more inference time than CONF-C, CONF-A, CONF-D, SF, and CNN, it has shown significantly lower error than these models. Our model also provides average and instantaneous RR simultaneously, but this functionality is missing in SF, CNN, CONF-A, CONF-D. Compared to the RN and CONF-B, the proposed model has shown lesser error, PC, and inference time than RN and CONF-B.

Noticeably, among the configurations that take raw signal as input CONF-A, CONF-B as given in Figure \ref{fig:new_model_config}, the multitasking model-based configuration CONF-B, shows lower error, but with a higher parameter count and inference time. Hence, the multitasking model provides lesser error irrespective of the type of input given. Between two multitasking configurations CONF-B and CONF-E, the CONF-E operates at the optimal time and PC as given in Table \ref{dl_model}, which may be attributed to the smaller respiration signals that CONF-E takes as input. On a broad scale, considering the optimal tradeoff between accuracy and parameter count, CONF-E has performed efficiently while taking an optimal time for execution. The instantaneous RR output from CONF-E is considered for further statistical analysis.

\begin{figure}[b]
    \centering
    \vspace{-1.2em}
    \includegraphics[width=1.0\linewidth, height= 0.5 \linewidth]{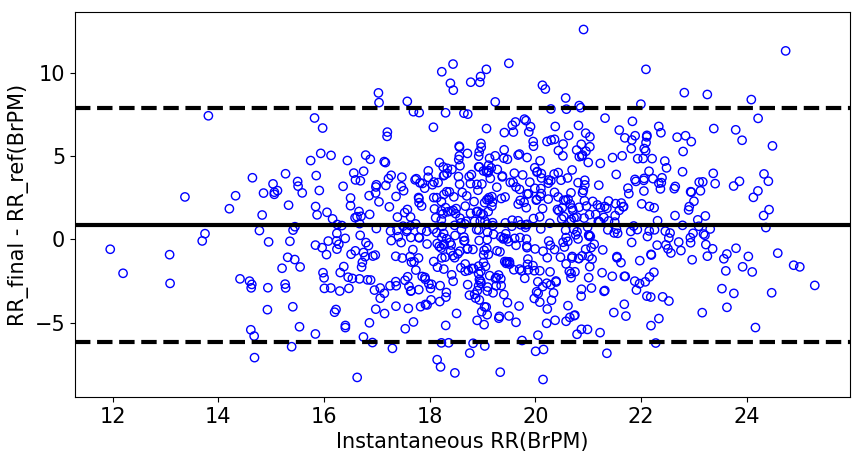}
    \caption{Bland Altman analysis between the final respiration rate and reference respiration rate.}
    \label{fig:bland altman}
    \vspace{-1.2em}
    
\end{figure}
\subsection{\textbf{Bland Altman Analysis}}
The box plot aids in understanding the errors concerning ground truth but, it does not account for the possible errors in estimating the ground truth RR itself. This aspect can be understood through the degree of agreement between the final RR output and reference RR obtained from the Bland Altman analysis shown in Figure \ref{fig:bland altman}. The limits of agreement obtained from the analysis are \textbf{7.90 BrPM} and \textbf{-6.14 BrPM} having a mean bias of \textbf{0.88 BrPM}, with \textbf{95.21\%} of the points lieing within the limit of agreement. The presence of outliers in the plot accounts for the highly erroneous modality during certain activities.
Overall, the multitasking network provides reasonable accuracy for both average RR and instantaneous RR. Hence this network is a reliable choice to be used for measuring RR during point of care settings.

\section{Conclusion}
This study has proposed a multitasking DL-based network to estimate RR and the respiration signal from ECG and accelerometer waveforms during ambulatory activities. While being relatively parameter efficient and consuming optimal time compared to other DL configurations, it also gives the best instantaneous RR. We have also conducted an extensive statistical analysis of the model during various activities and demonstrated its efficacy.    

Despite the robust performance of the proposed network, it requires more parameters relatively. Our future work would involve reducing the model complexity without compromising accuracy through methods like Knowledge Distillation. Due to reduced parameter count and time complexity, we would test the model's efficacy in a real-time scenario.
 
\bibliographystyle{IEEEtran}
\bibliography{root}
\end{document}